\documentstyle[11pt]{article}
\newcommand{\be}{\begin{eqnarray}}

\newcommand{\ee}{\end{eqnarray}}

\def\toneth{{\textstyle{\frac{1}{3}}}}

\title{
	\begin{flushright}
	{\normalsize TPI--MINN--94--7/T \\
	NUC--MINN--94--2/T \\
        HEP--MINN--94--1242/T \\
	February 1994 \\}
	\end{flushright}
\bf  Green's Functions in the Color Field of a Large Nucleus}
\author{
	Larry McLerran and Raju Venugopalan \\
        {\small\it Theoretical Physics Institute} \\
        {\small\it  School of Physics and Astronomy} \\
 	{\small\it  University of Minnesota} \\
        {\small\it Minneapolis, MN 55455} \\
	 }
\date{}

\begin{document}

\maketitle

\begin{center}
{\bf Abstract}\\
\end{center}
We compute the Green's functions  for scalars, fermions and vectors in the
color field associated with the infinite momentum frame wavefunction of a large
nucleus.  Expectation values of this wavefunction can be computed by
integrating over random orientations of the valence quark charge density.  This
relates the Green's functions to correlation functions of a two dimensional,
ultraviolet finite, field theory.  We show how one can compute the sea quark
distribution functions, and explictly compute them in the kinematic range of
transverse momenta, $\alpha_s^2 \mu^2 << k_t^2 << \mu^2$, where $\mu^2$ is the
average color charge squared per unit area. When $m_{quark}^2 << \mu^2 \sim
A^{1/3}$, the sea quark contribution to the infinite momentum frame wave
function saturates at a value that is the same as that for massless sea quarks.

\vfill \eject

\section{Introduction}

 In two previous papers~\cite{ven1,ven2}, we argued that the quark and gluon
distribution functions for very large nuclei at small values of Bjorken $x$
were computable in a weakly coupled many body theory. We argued that when $x <<
A^{-1/3}$ and when a parameter proportional to the density of valence quarks
per unit area,
\be
	\mu^2 = 1.1\, A^{1/3}~ {\rm fm}^{-2} \,,
\ee
is large, the theory is governed by a weak coupling  constant $\alpha_s(\mu)$.
The valence quarks serve as sources of color charge and can be treated as
static sources along the light cone. They follow straight line trajectories
travelling at the speed of light.  We further argued that as long as we were
measuring parton distribution functions on transverse momentum scales which are
$k_t^2 << \mu^2$, the sources of valence charge could be treated classically.
The field is stochastic and we must average over the classical sources of
charge $\rho$ with a Gaussian weight
\be
\int [d\rho ]~\exp{\left( - {1 \over {2\mu^2}} \int d^2x_t~\rho^2
(x)\right)} \, .
\ee

To lowest order in $\alpha_s$, the theory can be reduced to computations of the
two dimensional Euclidean correlation functions of an ultraviolet finite gauge
theory.

These correlation functions are related to the solution of the classical
equations of motion in the presence of an external current which is localized
on the light cone,
\be
	J^\mu_a (x) = \delta^{\mu +} \delta (x^-) \rho_a (x^+,x_t) \, .
\label{curr}
\ee
This source corresponds to a sheet of charge in the two transverse spatial
dimensions propagating at the speed of light $x = t$. Here we use the light
cone variables
\be
	x^{\pm} = {1 \over \sqrt{2}} (x^0 \pm x^3) \, .
\ee
In the light cone gauge,  $A_- = - A^+ = 0$, a solution to the classical
equations of motion for the gauge fields is $A_+ = -A^- = 0$ with the
transverse components of the gauge field given by
\be
	A_j (x) = \theta(x^-) \alpha_j(x_t) \, .
\ee
Further, the field $\alpha_j$ may be inserted into the equations of motion to
show that
\be
	F_t = 0 \,,
\ee
where $F_t$ are the transverse components of the field strength tensor,
and
\be
	\nabla \cdot \alpha = g \rho \, .
\ee
These conditions are equivalent to $\alpha_j$ being a pure gauge transform of
the vacuum for a two dimensional Yang-Mills theory, with the gauge condition
being the above equation.  We may therefore write
\be
	\alpha_j = -{1 \over {ig}}~ U(x_t) \nabla_j U^\dagger (x_t) \, ,
\ee
where the equation which determines $U$ is
\be
	\vec{\nabla} \cdot (U \vec{\nabla} U^\dagger ) = -ig^2 \rho \, .
\ee
This solution has zero light cone Hamiltonian, $P^- = 0$.

Finally, the integration over  all color orientations of the external sheet of
charge must be performed.  This is equivalent to computing the expectation
values of $<\alpha_i(x_t) \alpha_j(y_t)>$ with the measure
\be
	\int [d\alpha]~ \exp\left\{-{1 \over {g^2 \mu^2}}\int d^2 x_t  (\nabla
\cdot \alpha )^2 \right\} \delta(F_t) \,{\rm det}(\nabla \cdot D) \, ,
\ee
where $D$ is the covariant derivative. This measure can also be expressed in
terms of the compact fields $U$ as
\be
	\int [dU] ~\exp\left( -{1 \over {g^4\mu^2}}\int d^2x_t {\rm Tr}
\left[\vec{\nabla} \cdot (U{1 \over i}\vec{\nabla} U^\dagger ) \right]^2
\right)
{\rm det}(\vec{\nabla} \cdot \vec{D} )
\label{fluct}
\ee
The above  form for the equations explicitly demonstrates a scaling behavior.
Namely,  the two dimensional expectation values of the variables $U$ are
functions of $g^2 \mu \,x_t$ or $g^2 \mu/k_t$. The dependence on $k^+$ of the
external field is particularly simple and has the form $1/k^+$.  Thus to lowest
order in $\alpha_s$, and to all orders in $\alpha_s^2 \mu^2$, the distribution
function for gluons is flat in rapidity, and has the above simple dependence on
transverse momentum scales.  When $k_t >> \alpha_s \mu$, we showed in our
previous papers that the distribution functions were simply
Wiezs\"acker--Williams distributions.

Although the limitation that the computations we employ are only valid for
$k_t^2 << \mu^2$ arose from requiring that we can treat the external sources
classically, it would seem that one might easily  extend the kinematic region
where our results hold to larger values of transverse momentum.  After all, in
this region, the gluon field is essentially the perturbative
Wiezs\"acker--Williams field, with the only essential modification that the
source is quantum mechanical.  In lowest order, the source squared averages to
the same values as for the classical result, and the result is the same.

The problem is that there will be big quantum corrections
to our lowest order result.  These will arise as factors involving
$\alpha_s \ln(k_t/\mu )$ and can be large.  This is a consequence
of the fact that the background field is becoming weak, of order
$A_{k}^2 \sim \alpha_s \mu^2 /k_t^2$ and when $k_t \sim \mu$, the background
field is of the order of the quantum corrections.

Eventually it may be possible to extend the range of validity of our method to
larger values of $k_t$.   Indeed, the corrections seem to be similar to
corrections to bremstrahlung radiation which are understood to some degree. At
this time we do not know how to make these corrections, and our results are
restricted to transverse momenta $k_t << \mu$.  This is a non-trivial
restriction because the bremstrahlung spectrum is hard, and as we shall see for
quarks with masses $m_{quark} >> \mu$, the dominant contribution to the sea
quark spectrum comes from this region.

The central issue we will address in this paper is that of computing the
Green's functions for scalars, fermions and vectors in the presence of the
above background field.  We will consider scalars and fermions in the
fundamental representation of the gauge group. The vectors will be the gluons.
We shall first compute the Green's function before averaging over all values of
the valence quark charge, and later average over all values to obtain our final
result.  We will get explicit expressions for the Green's functions in terms of
the background field, and will thus be able to determine the scaling properties
of the distribution functions associated with these Green's functions.

It is not too surprising that the Green's functions can be explicitly computed.
In the region $x^- <0$ and in the region $x^- > 0$, the background field is a
pure gauge.  Only the step function at $x^- = 0$ prevents the field from being
entirely a gauge transform of the vacuum configuration.

There are atleast two uses towards which the Green's function computation can
be
applied.  The first is to compute, to lowest order in $\alpha_s$, the
contribution of sea quarks to the wavefunction of a nucleus. This is of
interest for heavy quarks because the enhancement of the momentum scale arising
from the typically higher density of partons in a nucleus leads to a
correspondingly enhanced contribution of the strange and charm quarks to the
nuclear wavefunction.   In fact we will see that if $m_{quark} << \mu$, then
the contribution of sea quarks has saturated at a value which is the same as
that of the massless sea quarks.

The Green's functions may also be applied to determine the higher order
corrections in $\alpha_s$  to the quark and gluon distribution functions.  This
will be the subject of a later analysis, where we will explicitly compute the
first order corrections.  It will be necessary to understand the pattern of
such corrections if the Lipatov enhancement~\cite{lipatov} is to be understood.
Hopefully, all of this will be feasible.

For the problems we wish to study in this paper, it is useful to know the
relation between the Green's functions and the distribution functions for
quarks
and gluons. We explicitly derive this relationship in the second section.

In the third section, we compute, in the fundamental representation, the scalar
particle Green's function in the presence of the background field.  We show how
averaging over the different orientations of the sources of charge simplifies
the result.

In the fourth section, we generalize our results for the Green's
function to fermions in the fundamental representation and to gluons.

In the fifth section, we compute the sea quark contribution to the nuclear
wavefunction.  We compute the ratio of the contributions of light mass quarks
to glue. For heavy quarks, we compute the mass dependence of our results.
Unfortunately, for heavy quarks with $m_{quark} \ge \mu$, the
dominant contribution to the integrated spectrum comes for
values of $k_t >> \mu$.

We will summarize our results in the final section. In an appendix, we obtain
an expression for the quantum fluctuations in the constrained vector fields.
These fields, which are not dynamical fields, are nevertheless necessary for
computations of some pieces in the gluon Green's functions.

\section{Propagators and Distribution Functions}

We want to be able to relate the distribution functions for various species of
particles to the propagators for these particles. Let us first consider the
example of a scalar field in the fundamental representation of the gauge group.
The  scalar field  may be written in terms of creation and annihilation
operators as
\be
	\phi^\alpha (x^+,\vec{x}) = \int_{k^+ > 0} {{d^3k} \over
{\sqrt{2k^+} (2\pi )^3}} \left\{ e^{ikx} a^\alpha (x^+,\vec{k})
+e^{-ikx} b^{\alpha \dagger} (x^+,\vec{k}) \right\} \, .
\ee
In this equation, $\vec{k}$ denotes the set $k^+, \vec{k}_t$. The equal (light
cone) time commutation relations for the $a$ and $b$ fields are
\be
	[a^\alpha (x^+,\vec{k} ) ,a^{\beta\dagger}(x^+,\vec{q} ) ] = (2\pi )^3
\delta^{(3)} (\vec{k}-\vec{q})\delta^{\alpha\beta}\, ,
\ee
and
\be
	[b^\alpha (x^+,\vec{k} ) ,b^{\beta\dagger}(x^+,\vec{q} ) ] = (2\pi )^3
\delta^{(3)} (\vec{k}-\vec{q})\delta^{\alpha\beta} \, ,
\ee
with all other commutators vanishing. The operator $a^\dagger $ creates a
scalar particle and the operator $b^\dagger $ creates its anti--particle charge
conjugated partner. If we define
\be
	b^{\alpha \dagger} (x^+,\vec{k} ) = a^\alpha (x^+,\vec{-k})
\ee
we then have
\be
	\phi^\alpha (x) = \int {{d^3k} \over {\sqrt{2\mid k^+ \mid }
(2\pi)^3 }} ~e^{ikx} a^\alpha (x^+,\vec{k}) \, .
\ee

The distribution function for the scalars is
\be
	{{dN} \over {d^3k}} ={1\over {(2\pi)^3}} \sum_\alpha <a^{\alpha
\dagger} (x^+,\vec{k})
a^{\alpha } (x^+,\vec{k}) +b^{\alpha  \dagger}(x^+,\vec{k})
b^{\alpha } (x^+,\vec{k}) > \, .
\ee
The sum in the above equation goes over both particles and anti--particles.
For the systems we consider, the sum will be symmetric under interchange of
particles and anti--particles.  We therefore have
\be
	{{dN} \over {d^3k}} =  {2\over {(2\pi)^3}}\sum_\alpha <a^{\alpha
\dagger} (x^+,\vec{k})
a^{\alpha } (x^+,\vec{k})  > \, .
\ee

Now on the other hand, we also have that
\be
	a^\alpha (x^+,\vec{k} ) = \sqrt{2 \mid k^+ \mid  }
\int d^3x ~e^{-ikx} \phi^\alpha (x)  \, ,
\ee
so that
\be
	{{dN} \over {d^3k}} =  2 i~{(2 k^+)  \over {(2\pi)^3}}~
\sum_\alpha D^{\alpha \alpha }
(x^+,\vec{k}, x^+,\vec{k} ) \, .
\ee
The propagator in the above equation is defined as
\be
	D^{\alpha \beta } (x,y) =  -i< \phi^\alpha (x) \overline
\phi^{~\beta} (y) >\, ,
\ee
with
\be
	D(x^+,\vec{k}, y^+,\vec{q} ) = \int d^3x d^3y ~e^{-ikx+iqy} D(x,y) \, .
\ee
Finally, we can write the distribution function as the following integral over
the fully Fourier transformed propagator,
\be
	{{dN} \over {d^3k}} = 2i~ {{2k^+} \over {(2\pi )^3}}
{}~\sum_\alpha~
\int {{dk^-} \over {2\pi} } {{dq^-} \over {2\pi }}~
D^{\alpha \alpha } (k^-, \vec{k},q^-,\vec{k} ) \, .
\label{scadist}
\ee

In general, our propagator will have the structure
\be
	D^{\alpha \beta } (k,q) = \delta^{\alpha \beta }
(2\pi )^3 \delta (k^- - q^-) \delta^{(2)} (\vec{k}_t - \vec{q}_t )
\Delta (k^+,q^+,k^-,\vec{k}_t ) \, .
\ee
This form follows because, after the averaging over sources, the propagator
must obey translational invariance in the transverse spatial directions and
color invariance. The external field does not depend on the light cone time
$x^+$--hence one also obtains a delta function of $k^- -q^-$.  Since the
background field depends on $x^-$ and $y^-$  even after summing over the
charges of the valence quarks, there is a  non--trivial dependence of the
propagator on both $k^+$ and $q^+$.

Combining terms together, we see finally that
\be
	{{dN^{scalar}} \over {d^3k}} = {{2i n_{scalar} (2k^+) } \over
{(2\pi )^3}}
{}~\pi R^2 \int {{dk^-}Ê\over {2\pi }}  \Delta^{scalar} (k^+,k^+,k^-,\vec{k}_t
) \, .
\ee

In exactly the same way, we derive for gluons
\be
	{{dN^{gluon}} \over {d^3k}} = {{i n_{gluon} (2k^+) } \over {(2\pi
)^3}}
{}~\pi R^2 \int {{dk^-} \over {2\pi }}  \sum _i \Delta^{gluon}_{ii}
(k^+,k^+,k^-,\vec{k}_t ) \, ,
\ee
where the sum over $i$ is a sum over transverse gluon polarizations.

For fermions we get
\be
	{{dN^{fermion}} \over {d^3k}} = {{2i n_{fermion} (2k^+) } \over
{(2\pi )^3}}
{}~\pi R^2 \int {{dk^-}\over {2\pi }}  \sum_s \Delta^{fermion}_{ss}
(k^+,k^+,k^-,\vec{k}_t ) \, .
\ee
Here the sum over $s$ is a sum over fermion spin degreees of freedom. In the
above, $n_{scalar}$, $n_{gluon}$ and $n_{fermion}$ are the respective color
degeneracies of the scalars, gluons and fermions. In the fundamental
representation this factor is $N_c$ and in the adjoint representation it is
$N_c^2-1$. A factor of two for the degeneracy of particles and anti--particles
is included in the above expressions. Further, the sum over spins and
polarizations will reproduce the spin degeneracy factors.

\section{The Propagator for Scalar Bosons}

We shall now compute the propagator for a scalar field in the
fundamental representation of the gauge group propagating in the
background gauge field
\be
	A^a_+ & =  & 0  \nonumber \\
         A^a_- & = & = 0  \nonumber \\
         \tau \cdot A_i & = & \theta (x^-) \alpha_i (x_t) \, ,
\ee
where
\be
	\alpha_i(x_t) = -{1 \over {ig}} U(x_t) \nabla_i U^\dagger (x_t) \, .
\ee
To do this, we first solve the Klein--Gordon equation
\be
	\left\{ -(\vec{\nabla}_t - ig \vec{\alpha} )^2 +
2\partial^+\partial^- + M^2 \right\} \phi_\lambda (x) = \lambda
\phi_\lambda (x)
\ee
for the scalar field $\phi$ and the corresponding equation for $\overline \phi
(x)$, the complex conjugate of the scalar field.

We will soon see that an eigenstate of the above equation is labeled
by its four momentum $p$ and its color label $s$ for a color spinor
with index $\beta $.  The Green's function is given in terms of the
above  eigenfunctions as
\be
	G^{\beta \delta} (x,x') = \int {{d\lambda } \over {\lambda -
i\epsilon}} \int {{d^4p} \over {(2\pi )^4}}~  \delta(\lambda - p^2 - M^2
)~ \sum_s \phi^{\beta s}_{\lambda p} (x) \overline \phi^{~\delta
s}_{\lambda p} (x^\prime) \, .
\label{gengre}
\ee
The solutions are normalized so that
\be
	\int d^4x~  \overline \phi^{~s^\prime}_{\lambda^\prime p^\prime
} (x) \phi^s_{\lambda p} (x)~ = ~\delta^{ss^\prime }
\delta^{(4)} (p^\prime - p) \, .
\ee

For $x^- < 0$, the external field vanishes.  The solutions are
therefore plane waves
\be
	\phi^{\alpha s}_{\lambda p} (x) = \exp\left( ip_t\,x_t -ip^-x^+
-{{p_t^2 + M^2 -\lambda} \over {2p^-}}x^- \right) u^\alpha_s \, .
\ee
Here $u$ is the elementary color spinor, such that
\be
	u^\dagger_{s^\prime} u_s = \delta_{s^\prime s} \, .
\ee

For $x^- > 0$, the field is a gauge transformation of the vacuum field
configuration.  The solutions for fixed $p$ are therefore
\be
	\phi^{\alpha s}_{\lambda p} (x) = \left( U(x_t) u_s \right)^\alpha
\exp\left( ip_t\,x_t -ip^-x^+
-{{p_t^2 + M^2 -\lambda} \over {2p^-}}x^- \right)  \, .
\ee

We must now construct the solution to the equations of motion which is
continuous across the discontinuity in $x^-$. The solution is
\be
	\phi^{\alpha s}_{\lambda p} (x)& = & e^{ip_tx_t-ip^-x^+}\bigg\{
\theta (-x^-) u^\alpha_s
\exp\big(-i{{p_t^2 + M^2 -\lambda} \over {2p^-}}\,x^- \big)  \nonumber \\
 &+ & \theta (x^-) \int {{d^2 q_t} ~\over {(2\pi )^2}}~ e^{iq_tx_t}
\exp\big(-i {{(p_t+q_t)^2 + M^2 -\lambda} \over {2p^-}}\,x^-\big) \nonumber \\
&\times &
 \left( U(x_t)
V^\dagger (q_t) u_s \right)^\alpha \bigg\} \, ,
\ee
where we have defined
\be
	V^\dagger (p_t) = \int d^2 x_t~ e^{-ip_tx_t} U^\dagger (x_t) \, .
\ee
The above solution for $\phi $ is properly normalized as may be
easily verified.  Upon defining
\be
	p^+ = {{p_t^2 + M^2 -\lambda } \over {2p^-}} \, ,
\ee
$\phi$ may alternatively be written as
\be
	\phi^{\alpha s}_{\lambda p} (x)& = & e^{ipx}
\bigg\{\theta (-x^-) u^\alpha_s \nonumber \\
&+ &
\theta (x^-) \int {{d^2 q_t} \over {(2\pi )^2}}~ e^{iq_tx_t}~
\exp\left(-i {{(p_t+q_t)^2 - p_t^2} \over {2p^-}}x^- \right)  \nonumber \\
&\times & \left( U(x_t)
V^\dagger (q_t) u_s \right)^\alpha\bigg\} \, .
\ee

With the above solutions in hand, we can now construct the Green's function for
the scalar field in an arbitrary background field of the type described above.
It is important to note that this expression makes no assumption about the
color averaging over the color labels of the external sources corresponding to
the valence quarks.  This averaging will be done later.  (The expression we
quote before averaging is the quantity which will be useful in loop graph
computations.)  The result of a considerable amount of straightforward but
tedious algebra is
\be
G^{\alpha \beta} (x,y) & = & \theta (-x^-) \theta (-y^-)
G_0^{\alpha \beta} (x-y) \nonumber \\
&+ & \theta (x^-) \theta (y^-) \left(U(x_t) G_0 (x-y)
U^\dagger (y_t)\right)^{\alpha \beta} \nonumber \\
&+ &
 \int {{d^4p} \over {(2\pi )^4}} {1 \over {p^2 + M^2 -i\epsilon}}
e^{ip(x-y)} \int {{d^2q_t} \over {(2\pi )^2}} d^2z_t~\bigg\{  \theta (x^-)
\theta (-y^-) \nonumber \\
&\times &
e^{i(p_t-q_t)(z_t - x_t)}
e^{\left[ -i {{(q_t^2 -p_t^2)} \over {2p^-}}x^- \right]}
\left(U(x_t) U^\dagger (z_t)\right)^{\alpha\beta} \nonumber \\
&+ &
\theta (-x^-) \theta (y^-) e^{-i(p_t-q_t)(z_t- y_t)}
\exp{\left[i {{(q_t^2 -p_t^2)} \over {2p^-}}y^- \right]}
\left(U(z_t) U^\dagger (y_t)\right)^{\alpha\beta} \bigg\}\, ,\nonumber \\
& &
\ee
where the free particle Green's function
\be
G_0^{\alpha\beta}(x-y)=\delta^{\alpha\beta}\int {d^4p\over {(2\pi)^4}}
{e^{ip\cdot (x-y)}\over{p^2+M^2-i\epsilon}} \, .
\ee
Now if we average the above Green's function over all possible
values of the color labels corresponding to the valence quarks, the
result simplifies even further.  Defining
\be
	<\left(U (x_t) U^\dagger (y_t)\right)^{\alpha\beta} > =  \delta^{\alpha
\beta}~\Gamma (x_t - y_t) \, ,
\ee
we see that
\be
	\Gamma (0) = 1 \, ,
\ee
which follows from the unitarity of the matrices $U$.
Defining the Fourier transform of $\Gamma$
\be
	\gamma (p_t) = \int d^2x_t~ e^{-ip_tx_t} \Gamma(x_t) \, ,
\ee
we have the sum rule
\be
	\int {{d^2p_t} \over {(2\pi )^2}}~ \gamma (p_t) = 1 \, .
\label{rule}
\ee

This expression can now be inserted into the definition of the propagator.  The
result can be expressed most simply in terms of the Fourier transform of the
Green's function--the propagator $D(p,q)$.  Letting $\delta D(p,q) = D(p,q) -
D_0(p,q)$ where $D_0$ is the free propagator, and writing
\be
	\delta D^{\alpha \beta}
(p,q) =  i\delta^{\alpha \beta} (2\pi )^3 \delta (p^- - q^-) \delta^{(2)}
(\vec{p}_t - \vec{q}_t) \delta \Delta (p,q)
\ee
it is straightforward to compute $\delta \Delta$. As a result of considerable
algebra and using the above sum rule for the integral of $\gamma (p_t)$, we
find finally that
\be
	\delta \Delta (p,q)&  = &
 \theta (p^-) \left( {1 \over {p^+-q^+ + i\epsilon}} - 2p^-
\Delta_0(q) \right)\nonumber \\
& &  \times \int {{d^2l_t} \over {(2\pi)^2}}~ \Delta_0(p -
l_t) \left(\gamma (l_t) - (2\pi )^2 \delta^{(2)} (l_t) \right) \nonumber \\
& &
+ \theta (-p^-) \left( {1 \over {p^+-q^+ + i\epsilon}} - 2p^-
\Delta_0(p) \right) \nonumber \\
& &  \times \int {{d^2l_t} \over {(2\pi)^2}}~ \Delta_0(q -
l_t) \left(\gamma (l_t) - (2\pi )^2 \delta^{(2)} (l_t) \right) \, .
\label{scaprop}
\ee
In the above, $\Delta_0$ represents the usual scalar propagator:
\be
\Delta_0 (p)={1\over p^2+M^2-i\epsilon} \, .
\ee

\section{Propagator for Fermions and Gluons}

It is easy to show that each component of the free fermion wavefunction obeys
the Klein Gordon equation. Hence the Green's function defined in Eq.
(\ref{gengre}) for the fermions is simply related to the scalar Green's
function by a relative normalization factor of $2k^+$. This relation also holds
for their Fourier transforms--the propagators. Explicitly,
\be
\delta\Delta (p,q)^{ferm}_{ss}=2p^+ \delta\Delta (p,q) \, ,
\label{fermpro}
\ee
where $\delta\Delta (p,q)$ is given in Eq. (\ref{scaprop}).

The solutions of the small fluctuation equations for the gluons in the
background field are plane waves in the adjoint representation for $x^-<0$ and
gauge transforms of plane waves in the adjoint representation for $x^->0$.
Defining $p^+=(p_t^2-\lambda)/2p^-$, and matching the fields across the
discontinuity at $x^-=0$, we obtain for the transverse components of the gauge
field the relation
\be
A_t^{\alpha\beta}&=& e^{ipx}\eta_t\bigg\{\theta(-x^-) \tau^{\alpha\beta}+
\theta(x^-) \int {{d^2 q_t} \over {(2\pi)^2} }e^{iq_t x_t} \nonumber \\
&\times &
\exp{-i\left[{{2p_tq_t+q_t^2}\over 2p^-}x^-\right]}\int d^2 \bar{x}_t
e^{-iq_t\bar{x}_t}\left(U(x_t) U^{\dagger}(\bar{x}_t)\tau U(\bar{x}_t)
U^\dagger
(x_t)\right)^{\alpha\beta}\bigg\} \, ,\nonumber \\
\label{tvec}
\ee
where $\eta$ is a unit vector and the $\tau$'s correspond to the usual SU(3)
matrices. The computation of the Green's functions for the vector case is
exactly analogous to that of the scalar Green's functions outlined in the
previous section. To obtain our final result though we have to make use of the
Fierz identity
\be
(\tau)^{\alpha\delta}(\tau)^{\gamma\beta}={1\over 2}\left(\delta^{\alpha\beta}
\delta^{\gamma\delta}-{1\over
3}\delta^{\alpha\delta}\delta^{\gamma\beta}\right)
\, .
\ee
Our result for the Green's function is then
\be
G^{\alpha\beta;\alpha^\prime \beta^\prime}(x,y)&=& \int {{d^4p}\over
{(2\pi)^4}}
{{e^{ip(x-y)}}\over {p^2-i\epsilon}}\Bigg\{\theta(-x^-)\theta(-y^-)
(\tau)^{\alpha\beta}(\tau)^{\alpha^\prime\beta^\prime}\nonumber \\
&+& \theta(x^-)\theta(y^-) \left(U(x_t)\tau U^\dagger (x_t)\right)^{
\alpha\beta}\left(U(y_t)\tau U^\dagger (y_t)\right)^{
\alpha^\prime\beta^\prime}\nonumber \\
&+& \int {{d^2q_t}\over {(2\pi)^2}}d^2\bar{x}_t\bigg[\theta(x^-)\theta(-y^-)
e^{i(p_t-q_t)(\bar{x}_t-x_t)} e^{-i(q_t^2-p_t^2) x^-/2p^-}\nonumber \\
&\times  &\left(U(x_t)\tau
U^\dagger (x_t)\right)^{ \alpha\beta}\left(U(\bar{x}_t)\tau U^\dagger (\bar{x}_
t)\right)^{\alpha^\prime\beta^\prime} \nonumber \\
&+&\theta(-x^-)\theta(y^-)
e^{i(p_t-q_t)(y_t-\bar{x}_t)} e^{i(q_t^2-p_t^2) y^-/2p^-}\nonumber \\
&\times &\left(U(\bar{x}_t)\tau
U^\dagger (\bar{x}_ t)\right)^{\alpha\beta} \left(U(y_t)\tau U^\dagger
(y_t)\right)^{ \alpha^\prime\beta^\prime}\bigg]\Bigg\}\, .
\ee

Again, as in the case of scalars, we define the expectation value
\be
\langle\left(U(x_t)\tau U^ \dagger
(x_t)\right)^{\alpha\beta}\left(U(y_t)\tau U^ \dagger
(y_t)\right)^{\alpha^\prime\beta^\prime}\rangle
= {1\over
2}\left(\delta^{\alpha\beta^\prime}\delta^{\alpha^\prime\beta}-{1\over
3}\delta^{\alpha\beta}\delta^{
\alpha^\prime\beta^\prime}\right)\Gamma(x_t-y_t) \, .\nonumber \\
\ee

The rest of the discussion is identical to that in the previous section and the
change in the gluon propagator $\delta D^{\alpha \beta ;\alpha^\prime \beta^
\prime} (p,q)$ may be expressed as
\be
	\delta D^{\alpha \beta ;\alpha^\prime \beta^\prime}
(p,q) = {1\over 2} i\left(\delta^{\alpha\beta^\prime} \delta^{\alpha^\prime
\beta}-\toneth
\delta^{\alpha\beta}\delta^{\alpha^\prime\beta^\prime}\right)   (2\pi )^3
\delta (p^- -
q^-) \delta^{(2)} (\vec{p}_t - \vec{q}_t) \delta \Delta (p,q) \, ,\nonumber \\
\ee
where $\delta\Delta (p,q)$ has a form identical to the result obtained in
Eq. (\ref{scaprop}).

\section{Sea Quark Distribution Functions}

Now that we have computed the fermion propagator in the classical background
field in the previous section, we are in a position to calculate, to lowest
order in $\alpha_s$, the sea quark distributions in this background field. The
relation between propagators and the corresponding distribution functions has
been discussed in section 2. However, due to the singular nature of the
propagators in the background field, the actual computation of the
distributions is somewhat subtle. This computation will be outlined below. In
the rest of this section, we compute the ratio of sea quarks to glue in the
background field and study the mass dependence of our results--whether one
obtains an enhanced contribution from strange and charm quarks to the nuclear
wavefunction.

{}From Eq. (\ref{scadist}) and Eq. (\ref{fermpro}), we can show that the sea
quark distribution function can be written in terms of the scalar propagator
as
\be
 {1\over {\pi R^2}}{{dN^{ferm}} \over {d^3k}} = {4 N_c (2k^+) \over
{(2\pi)^3}}i
\int d{k^+}^\prime d{k^-}^\prime \delta D(k_t,{k^-}^\prime ,k^+;{k^+}^\prime )
\delta ({k^+}^\prime -k^+)
 \, .
\label{sea1}
\ee
Note that we have implicitly included a factor of two from the two light cone
spin degrees of freedom in the above. Typically, the above equation would
reduce to
\be
 {1\over {\pi R^2}}{{dN^{ferm}} \over {d^3k}} = {4 N_c (2k^+) \over
{(2\pi)^3}}i
\int {dk^-\over {2\pi}} \delta D(k_t,{k^-} ,k^+;{k^+})
 \, .
\ee
However, this cannot be done in this case due to the singularity in $\delta
D(k_t,{k^-}^\prime ,k^+;{k^+}^\prime )$ as ${k^+}^\prime \rightarrow k^+$ (see
Eq. (\ref{scaprop})). The limit must be taken only at the end--after performing
the integrals in Eq. (\ref{sea1}).

Substituting Eq. (\ref{scaprop}) in Eq. (\ref{sea1}), we obtain
\be
 {1\over {\pi R^2}}{{dN^{ferm}} \over {d^3k}} &=& {4 N_c (2k^+) \over
{(2\pi)^3}}
\lim_{{k^+}^\prime \rightarrow k^+}\int {d^2p_t \over {(2\pi )^2}} \left(
(2\pi )^2\delta^{(2)} (p_t) - \gamma (p_t)\right)\nonumber \\
&\times & \Bigg\{ \int_{-\infty}^{\infty} {d{k^-}^\prime \over {2\pi}}
(-2{k^-}^\prime ) \Delta_0 (k)\Delta_0 (k^\prime + p_t)\nonumber \\
&+& \left[\int_{-\infty}^0 {d{k^-}^\prime \over {2\pi}} {\Delta_0 (k+p_t)\over
 {{k^+}^\prime -k^+ + i\epsilon}} + \int_0^{\infty} {d{k^-}^\prime \over
{2\pi}} {\Delta_0 (k^\prime +p_t)\over  {{k^+}^\prime -k^+ + i\epsilon}}
\right]\Bigg\}
\, .
\label{sea2}
\ee
The first term within the curly brackets vanishes because both poles in the
contour integral lie on the same side of the contour. The remaining two
integrals are apparently logarithmically divergent. It can also be shown that
the pieces singular in the limit ${k^+}^\prime \rightarrow k^+$ either vanish
or cancel out. The sum of the two integrals in the brackets may then be written
as
\be
\lim_{Z\rightarrow \infty} {1\over 2}{{\left(\ln \left[ (k_t+p_t)^2+
M^2\right]-\ln Z \right)}\over {k^+}^2} \, ,
\ee
where $Z$ is a constant corresponding to the upper limit of the divergent
integrals.

Substituting the above result in Eq. (\ref{sea2}), we obtain
\be
 {1\over {\pi R^2}}{{dN^{ferm}} \over {d^3k}} &=& 4 {N_c (2k^+) \over
{(2\pi )^4}}
\int {d^2p_t \over {(2\pi )^2}} \left(\gamma (p_t)-(2\pi )^2\delta^{(2)}
(p_t)\right) \nonumber \\
&\times & {1\over 2}\lim_{Z\rightarrow \infty}{{\left(\ln \left[
(k_t+p_t)^2+ M^2\right]-\ln Z \right)}\over {k^+}^2} \, .
\ee
However, the $Z$--dependent piece of the above integral vanishes due to the
sum rule in Eq. (\ref{rule}),
\be
\int {d^2p_t \over {(2\pi )^2}} \left(\gamma (p_t)-(2\pi )^2\delta^{(2)}
(p_t)\right)=0 \, .
\ee

Our result for the sea quark distribution can therefore be written as
\be
 {1\over {\pi R^2}}{{dN^{ferm}} \over {d^3k}} &=& { 4 N_c\over {(2 \pi )^4
k^+}}
\int {d^2 p_t\over {(2\pi)^2}}
\gamma (p_t) \ln \left[{{(k_t+p_t)^2+M^2}\over {k_t^2+M^2}} \right]  \, .
\ee
A key feature of the above result is that the sea quarks demonstrate the same
$1/x$ scaling behaviour as that obeyed by the gluons.

The sea quark distribution in rapidity per unit rapidity per unit transverse
area can be represented as
\be
 {1\over {\pi R^2}} {dN\over dy}={ 4 N_c \over {(2 \pi )^4 }}
 \int {d^2 p_t\over {(2\pi)^2}}
\gamma (p_t) \int d^2 k_t \ln \left[{{(k_t+p_t)^2+M^2}\over {k_t^2+M^2}}
\right] \, .
\ee
It turns out that the integration over $k_t$ can be performed analytically and
one obtains the simple result $\pi p_t^2$.   This result implicitly assumes
that the range of $k_t$ and quark mass is $<< \mu$.  If otherwise, for reasons
stated in the introduction, we expect large contributions due to radiative
corrections.  The stated range of transverse momentum is the dominant one if we
require that the mass be $m_{quark} << \mu$. This is because this range of
the sea quark transverse momentum corresponds to the range $p_t \le \mu$ of the
gluon distribution--the contribution from the large $k_t \ge m_{quark}$ is
suppressed in this range of the gluon transverse momentum.

Now, in Ref.~\cite{ven2}, we argued that the dominant contribution from the
averaging over the color charge distribution (see Eq. (\ref{fluct})) came from
the range $\alpha_s^2\mu^2 << p_t^2$. In this range, the leading order
contribution to $\gamma(p_t)$ from the color averaging is simply
\be
\gamma (p_t)= {{2 (4\pi)^2 (N_c^2-1)}\over {2 N_c}} {\alpha_s^2\mu^2\over
{p_t^4}} \, . \ee

Therefore, in the dominant range of integration, $\alpha_s \mu << p_t << \mu$,
we obtain the astonishingly simple result for the sea quark rapidity
distribution
\be
{1\over {\pi R^2}}{dN_{sea}\over {dy}}= {{2(N_c^2-1)} \over {\pi^2}}
\mu^2 \alpha_s^2 \ln ({1\over {\alpha_s}}) \, .
\ee
This result is the main result of our paper. Recall that the lowest order
result for the gluon distribution arising from the same range of momenta was
just the Weizacker--Williams result scaled by $\mu^2$:
\be
{1\over {\pi R^2}} {dN_{glue}\over dy}= {2 \alpha_s \ln({1\over \alpha_s})
\mu^2 (N_c^2-1)\over \pi} \, .
\ee
The ratio of intrinsic quarks to glue in the nuclear wave function is therefore
suppressed by a factor $\alpha_s/\pi$. Note also that the sea quark
distribution saturates--it is not dependent on the sea quark mass. This result
is true for all quark masses which are $m_{quark} << \mu$. The dominant
contribution for heavier quarks comes from large transverse momentum where our
results cannot yet be extended.

\section{Summary}

We have given in this paper explicit expressions for the quark and gluon
propagators in the background field generated by a heavy nucleus in the
infinite momentum frame.  These expressions will allow us to compute the first
radiative corrections to the distribution functions. If there is a Lipatov
enhancement, it will appear in the first order radiative corrections. If such
an enhancement does occur, once understood,  a technique must be developed for
including its effects to all orders.  This issue will be the subject of  later
analyses.

We have estimated the contributions of light to intermediate mass sea quarks to
the distribution functions in the region  of momentum where our results are
reliable. To get a reliable estimate for larger masses, one must understand the
kinematic range when $k_t \ge \mu$.  This region is basically a weakly coupled
region, but is not yet understood within our framework. The region where
the distribution function is cutoff, $k_t \sim \alpha_s \mu$, must also be
properly understood if one wants to go beyond an approximation which is
accurate to leading logarithm of $\alpha_s$.  Finally, the modifications due to
a possible Lipatov enhancement must be understood.

There is also some hope that the result for heavy nuclei might be
extended to hadrons. This would occur if there is a Lipatov
enhancement, since in this case the typical parton separation
might become smaller than the hadronic size.  We have made the first step
toward computing this possible enhancement in the context of this theory.

There is finally the issue of how to relate these computations to experimental
measurement of structure functions.  This can only be done by studying the
dependence on the $Q^2$ of the probe by using an Altarelli-Parisi analysis.
There are two important weak coupling regions.  In the first, $Q^2 >> \mu^2$,
in which case, presumably, the ordinary Altarelli--Parisi analysis goes
through. The other region is $\mu^2 >> Q^2 >> \Lambda^2_{QCD}$.  In this region
the coupling is weak, but the intrinsic transverse momentum of the quarks and
glue is important. It is this region which must be studied carefully since this
region provides the greatest potential for measuring the intrinsic properties
of the hadronic wavefunction.

\section{Acknowledgments}

One of us (R.V) would like to thank Paolo Provero and Dietrich B\"odeker for
useful discussions. We acknowledge support under DOE High Energy
DE-AC02-83ER40105 and DOE Nuclear DE-FG02-87ER-40328.

\section{Appendix A}

We have derived in this paper expressions for the Green's functions for the
scalar, vector and fermion {\it independent} fields. One can also compute
Green's functions for the {\it dependent} fields by making use of the
constraint
conditions on the light cone. These constraints are obtained from the equations
of motion on the light cone (see the discussion in Ref.~\cite{ven1}).

In this brief appendix, we obtain an expression for the dependent vector field
$A^-$ in terms of the transverse vector field $A_t$. The Green's function for
the constrained fields may then be obtained following a procedure identical to
that followed in Section 4.

The general constraint condition for the $A^-$ vector field is obtained from
the equations of motion to be
\be
-\partial_{-}^2 A^-= J^+ + D_t E^t \, .
\ee
In the above equation, $J^+$ is the component of the light cone current defined
by Eq.~(\ref{curr}), $D_t$ are the transverse components of the covariant
derivative and $E^t=\partial_{-} A^t$ are the transverse electric fields.

If we define $A^{\mu}=A_{cl}^{\mu} + A_{qu}^{\mu}$, then keeping
only terms up to order $O(A_{qu}^{\mu})$, then
\be
-\partial_{-}^2 A_{qu}^-=D_t^{(0)}(\partial_{-} A_{cl}^t)-ig A_{qu,t}
(\partial_{-} A_{cl}^t) \, .
\label{constr}
\ee
Here $D_t^{(0)}=(\partial_t-ig A_{cl,t})$.

In Eq.~(\ref{tvec}), we had obtained an expression for
$A_{qu,t}^{\alpha\beta}$.
This expression can be re--written in the following compact form
\be
A_{qu,t}^{\alpha\beta}&=& e^{-ip^- x^+}\bigg\{\theta(-x^-) e^{-ip^+ x^-}
f_{1t}^{\alpha\beta}+ \theta(x^-) \int {{d^2 q_t} \over {(2\pi)^2}
}e^{i(q_t+p_t) x_t} \nonumber \\
&\times &\exp \left(-i\left[{{2p_tq_t+q_t^2}\over
2p^-}x^-\right]\right) f_{2t}^{\alpha\beta} (q_t)\bigg\} \, ,
\ee
where
\be
f_{1t}^{\alpha\beta} &=& \eta_t e^{ip_t x_t} \tau^{\alpha\beta} \, .\nonumber
\\
f_{2t}^{\alpha\beta} &=& \eta_t \int d^2\bar{x}_t e^{-q_t \bar{x}_t}
\left(U(x_t) U^{\dagger}(\bar{x}_t)\tau U(\bar{x}_t) U^\dagger
(x_t)\right)^{\alpha\beta}\, ,
\ee
where $\eta_t$ is a unit vector. Further, $A_{cl}^t=\theta (x^-)
\alpha_{cl}^t$.

To obtain an expression for $A_{+}$, we integrate both sides of Eq.~(\ref
{constr}). We choose the boundary condition $A^{-}(x^- = 0)=0$. This particular
boundary condition is convenient because it implies that the source current
$J^+$ has a constant value for all light cone times $x^+$. Though convenient,
this boundary condition is not sufficient to ensure that the fields vanish at
infinity as implied by the original constraint equation. We shall return to
this tricky issue in a later work.

Our final result for the quantum correction to the constrained field $A_{+}$
is
\be
(A_{+}^{qu})^{\alpha\beta}&=&  e^{-ip^+ x^- -ip^- x^+ + ip_t x_t}
\bigg\{-\theta(-x^-) (1-e^{ip^+ x^-}){p_t\over {p^+}}\eta_t\tau^{\alpha\beta}
\nonumber \\
&+&
\theta(x^-) {i\over {p^+}}\int {{d^2 q_t} \over {(2\pi)^2} }{e^{iq_t x_t}\over
{\left[1+ {
(2p_t q_t + q_t^2)\over {2p^+ p^-}}\right]}}
 \left(\exp (-ix^-{(2p_t q_t + q_t^2)\over {2p^-}}-
e^ip^+ x^-\right)
\nonumber \\
&\times &\left[ i(q_t + p_t)f_{2t}^{\alpha\beta}+\partial_t
f_{2t}^{\alpha\beta}- ig \alpha_t (x_t) f_{2t}^{\alpha\beta}\right]
\nonumber \\
&+& ig x^- e^{ip^+ x^-} \theta(x^-) [\eta_t \tau,\alpha_t]^{\alpha
\beta} \bigg\} \, .
\ee
Note that this solution for $A_{+}^{qu}$ is defined only upto a term $B x^-$,
where $B$ is an arbitrary function which does not depend on $x^-$. The Green's
functions for the constrained fields can now be computed in the manner
discussed in previous sections.

An essential feature of the above result is that it has terms infrared singular
in $p^+$. There is an extensive literature on how to regularize such
singularities which are endemic in light cone
quantization~\cite{Leib,Mandel,Basset,Soper}. A discussion of light cone
regularization schemes is outside the scope of this paper. However, this issue
will be important when one considers one loop corrections to our results and
the problem of regularization will be addressed more fully at that time.

\end{document}